\documentclass{aa}

\usepackage[]{graphics}
\usepackage[]{psfig}

\begin{document}

\thesaurus{03(11.03.04;11.05.1;11.09.2;11.19.4)}

\title{The central region of the Fornax cluster -- 
III. Dwarf galaxies, globular clusters, and cD halo -- are there interrelations?
}

\author {Michael Hilker \inst{1,2} \and Leopoldo Infante \inst{1} \and
Tom Richtler \inst{2}
}

\offprints {M.~Hilker}
\mail{mhilker@astro.uni-bonn.de}

\institute{
Departamento de Astronom\'\i a y Astrof\'\i sica, P.~Universidad Cat\'olica,
Casilla 104, Santiago 22, Chile
\and
Sternwarte der Universit\"at Bonn, Auf dem H\"ugel 71, 53121 Bonn, Germany
}

\date {Received --- / Accepted ---}

\titlerunning{The central region of the Fornax cluster -- dwarf galaxies, 
globular clusters, and cD halo}
\authorrunning{M.~Hilker et al.}
\maketitle

\begin{abstract}

In this paper we briefly review the properties of the dwarf galaxy population
at the core of the Fornax cluster, as well as the properties of the
extraordinary rich globular cluster system and the cD halo around the
central galaxy NGC~1399.
In turn, the question whether a scenario in which dwarf galaxies have
been accreted and dissolved in the cluster center can explain the observed
properties is addressed.
The possibility of accretion of a certain number of dwarf galaxies, the
stripping of their globular clusters and gas, and the formation of new
globular clusters from stripped infalling gas are discussed.
An increase in the specific frequency of the central globular cluster system
is only possible, if the infalling gas from stripped dwarfs formed globulars
very efficiently and/or accreted and stripped dwarf galaxies possessed
a rich globular cluster system themselves. In conclusion, we argue
that although the dwarf galaxy infall is a very attractive scenario to
explain a number of properties in the system there are probably other physical
mechanisms that also take place.


\end{abstract}

%%%%%%%%%%%%%%%%%%%%%%%%%%%%%%%%%%%%%%%%%%%%%%%%%%%%%%%%%%%%%%%%%%%%%%%%%%%
%%%%%%%%%%%%%%%%%%%%%%%%%%%%%%%%%%%%%%%%%%%%%%%%%%%%%%%%%%%%%%%%%%%%%%%%%%%
%%%%%%%%%%%%%%%%%%%%%%%%%%%%%%%%%%%%%%%%%%%%%%%%%%%%%%%%%%%%%%%%%%%%%%%%%%%

\section{Introduction}

The central regions of galaxy clusters are the places with the highest galaxy 
density
in the universe. Dwarf ellipticals (dE) are especially the most
strongly clustered types of galaxies in high-density environments (e.g. review 
by Ferguson \& Binggeli \cite{ferg94}, and references therein).
Several striking characteristics are seen in the center region of clusters:
(1) most central galaxies possess extraordinarily 
rich globular cluster systems (GCS) (see Harris \cite{harr91a}, Richtler 
\cite{rich95} and references 
therein), but see also apparent counter-examples (see Table 14 in
McLaughlin et al. \cite{mcla94b}); (2) there often
exists a cD galaxy in the center of clusters (e.g. Schombert \cite{scho88}). 
(3) different types of dwarf galaxies have 
different clustering properties (e.g. Vader \& Sandage \cite{vade91}); (4) in 
some cases the faint end slope of the dwarf galaxy luminosity function (LF)
seems to depend
on the cluster-centric distance (e.g. in Coma: Lobo et al. \cite{lobo}).
The question arises on whether these properties may
be related through the accretion of dwarf galaxies.

The answer to this question is most probably associated to the formation epoch 
of galaxy clusters. At that time, it is expected that galaxies were very
gas-rich and that interactions between galaxies were more frequent.
The number density of galaxies at that epoch in the central region must have 
been larger than today. Therefore, the initial population of dwarf
galaxies played an important role. The
favoured theoretical models of  galaxy cluster formation predict a
steep slope of the initial mass function towards the low-mass end (see
a more detailed discussion and references in Sect.~2.1). In contrast,
the faint end slope of the observed LF in nearby groups and clusters
are significantly flatter (see Ferguson \& Binggeli \cite{ferg94},
Trentham \cite{tren98}).  One possibility that would explain this
discrepancy is the accretion and  dissolution of dwarf galaxies in
cluster centers.  It is posible to understand the formation of  a rich
GC system and a cD halo from the infall of gas-poor and gas-rich
dwarfs into a dense cluster environment.
During the infall of gas-poor as well as gas-rich
dwarfs in a dense cluster  center environment several scenarios are
thinkable for forming a rich GCS and a cD halo (see Sect.~5).

Support for such a scenario from the observational side comes from L\'opez-Cruz
et al. (\cite{lope}) who compared the properties of 
clusters with and without a central luminous cD galaxy. They found that clusters
without a prominent cD galaxy tend to have a steep LF at the faint end and 
a high
fraction of late-type galaxies, and thus seem to be less evolved than
clusters with pronounced cD galaxies and relatively flat LF at the faint end.
They explain this finding by the disruption of dwarf galaxies.

In this study the attention is focused on the properties of the relatively
poor, compact, and evolved Fornax cluster, one of the best studied galaxy 
clusters in the local universe (e.g. Ferguson \cite{ferg89}, Ferguson \& 
Sandage \cite{ferg88}). 
Other nearby clusters are believed to be in different 
evolutionary states. Whereas Virgo (e.g. Sandage et al. \cite{sand85}, 
Ferguson \& Sandage \cite{ferg91}) is dominated by late type galaxies and is 
only half as dense in the center (numbers of galaxies per volume) as Fornax.
Centaurus (Jerjen \& Dressler \cite{jerj97b}, Stein et al. \cite {stei})
and Coma (e.g. Secker \& Harris \cite{seck97}) show substructures, 
indicative of a still ongoing dynamical evolution, like for example
cluster-cluster or cluster-group merging.

In the first two papers of this series (\cite{hilk99b}, \cite{hilk99a}, 
hereafter Paper~I and Paper~II)
we investigated the distribution of galaxies in central Fornax fields.
We found two compact objects that belong to the Fornax cluster and might be
candidates for isolated nuclei of stripped dwarf ellipticals.
However, very few new members were found compared to the study of
Ferguson (\cite{ferg89}). Thus, the spatial distribution and luminosity function
of dwarf galaxies in Fornax (Ferguson \& Sandage \cite{ferg88}) was confirmed.

In this paper we discuss the possibility, whether
the infall of dwarf galaxy into the cluster center may play an important role
in the enrichment of the central globular cluster system, especially the
increase of the globular cluster specific frequency $S_N$, as well as the
formation of the extended cD halo.

In the following section we give a compilation of the necessary background 
of our analysis.

%
%%%%%%%%%%%%%%%%%%%%%%%%%%%%%%%%%%%%%%%%%%%%%%%%%%%%%%%%%%%%%%%%%%%%%%%%%%%
%%%%%%%%%%%%%%%%%%%%%%%%%%%%%%%%%%%%%%%%%%%%%%%%%%%%%%%%%%%%%%%%%%%%%%%%%%%
%%%%%%%%%%%%%%%%%%%%%%%%%%%%%%%%%%%%%%%%%%%%%%%%%%%%%%%%%%%%%%%%%%%%%%%%%%%
%
\section{Dwarf galaxies in clusters}

\subsection{Theoretical background on the evolution of dwarf galaxies in 
clusters}

In their review about dwarf elliptical galaxies, Ferguson \& Binggeli 
(\cite{ferg94})
summarized the formation and evolutionary scenarios that are predicted
by theoretical models. It is generally accepted that galaxy formation
started from gaseous conditions in the early universe followed by the collapse 
of primordial density fluctuations, cooling of the gas and subsequent
star formation (e.g. White \& Frenk \cite{whit91}, Blanchard et al. \cite{blan},
Cole et al. \cite{cole94}, Kauffmann et al. \cite{kauf93}, Lacey et al.
\cite{lace}). In cold dark matter (CDM) dominated models the formation of
low-mass galaxies is favored, because for dwarf galaxy halos collapsing at
$z \simeq 3$--10 the cooling time is short compared to the free-fall time,
thus cooling should be very efficient, and accordingly many dwarfs will be 
formed.
A steep slope, $\alpha = -2$, of the initial mass function ($N(M)$d$M \propto
M^\alpha$) is predicted (e.g. Blanchard et al. \cite{blan}). In contrast, the
faint end slope of the observed luminosity functions in nearby clusters
are around $\alpha \simeq -1.3\pm0.4$ (see Ferguson \& Binggeli \cite{ferg94}).
This contradiction is the so-called
``overcooling problem'' (e.g. Cole \cite{cole91}). If the CDM model prediction
is correct, there must have been active some mechanisms
that either counteracted the cooling during the collapse of dwarfs or 
destroyed the numerous dwarfs after their formation. Plausible mechanisms that 
involve internal as well as external agents are summarized in the review 
by Ferguson \& Binggeli (\cite{ferg94}).

In the following, we focus our attention on the possibility that many dwarf
galaxies have merged with the central galaxy. For a CDM power spectrum in
an $\Omega = 1$ cosmology the epoch of dwarf galaxy formation is believed
to be also the epoch of rapid merging. Kauffmann et al. (\cite{kauf94}) included
the merging of satellite galaxies in their CDM models and found that most of the
observational data can be reproduced when adopting a merging timescale that
is a tenth of the tidal friction timescale, and when star formation is 
suppressed in low-circular-velocity halos until they are accreted into larger
systems. Further, efficient merging at all epochs results in a decrease of the
faint end slope of the LF compared to the initial predicted value of
$\alpha = -2$.

\subsection{Dwarf galaxies and cD halo}

Several authors have suggested that tidal disruption (total dissolution of 
the galaxy light) of galaxies in cluster centers as well as tidal stripping
(only outer parts are affected, a remnant survives) might be related to
the formation of cD halos (see references below).
The time of formation is being discussed. Most
authors assume that the stripping processes take place after the cluster 
collapse (e.g. Gallagher \& Ostriker \cite{gall72}, Richstone \cite{richs76}, 
Ostriker \& Hausman \cite{ostri}, Richstone \& Malumuth \cite{richs83}).
In contrast, Merritt (\cite{merr}) explained the general appearence of
cD halos as the result of dynamical processes during the cluster collapse.
In his scenario the accumulation of slowly-moving galaxies in the cluster core 
via dynamical friction only plays an important role for groups or clusters 
with small velocity dispersion $\sigma_v \leq 500$ km~s$^{-1}$ (Fornax: 
$\sigma_v \simeq 360$ km~s$^{-1}$).
White (\cite{whit87}) argued that, in the case of tidal disruption and
stripping, the distribution of 
stripped and disrupted material (diffuse light, dark matter, GCs) should be
more concentrated to the center than the relaxed galaxy distribution,
because galaxies closer to the center are more affected by disruption
processes than galaxies outside. In the case of Merritt's model, galaxies 
formed before
the collapse, stripping occured during the collapse, and finally the stripped
material is distributed in the same way as the galaxies through
collective relaxation.

\begin{table*}
\caption{\label{toverview} Overview of different parameters of Fornax 
components, as the population of dwarf elliptical and dwarf S0 galaxies,
the central GCS, the unresolved stellar light of NGC~1399 and its cD halo,
and the gas component in the central region of the cluster. The assumed 
distance moduludus is $(m - M)_0 = 31.3$ mag. The profile slope 
$\alpha$ is the exponent in the density law $\rho \propto r^\alpha$,
where $\rho$ is either the projected number density or the projected
surface brightness profile.}
\begin{flushleft}
\begin{minipage}{16cm}
\begin{tabular}{lccccc}
\hline\noalign{\smallskip}
cluster  & profile slope & velocity disp. & metallicity & $M_V$ & mass\\
component & $\alpha$ & $\sigma_v$ $[km/s]$ & $[Fe/H]$ & [mag] & [$M_{\sun}$]\\
\noalign{\smallskip}
\hline\noalign{\smallskip}
dE/dS0\footnote{nucleated and non-nucleated subsamples see Table~\ref{tslopes}} 
& $-1.0\pm0.2$ & 410 to 490 & $-1.5$ to $-0.8$ & $-20.8\pm0.2$ &\\
dE/dS0 ($r < 0\fdg7$) & $-0.2\pm0.1$ & &  & $-19.3\pm0.4$ &\\
dE/dS0 ($r > 0\fdg7$) & $-1.5\pm0.3$ & &  & $-20.4\pm0.3$ &\\[0.5ex]
central GCS & $-1.5\pm0.2$ & $373\pm35$ & $-1.8$ to 0.3  & $-17.6\pm0.2$ &\\
GCS (blue pop.) & $\simeq -1.0\pm0.2$\footnote{see Fig.~11 in Forbes et al. 
(\cite{forb97})} & $362\pm104$ & $-1.8$ to $-0.8$  & &\\
GCS (red pop.) & $\simeq -1.7\pm0.2^b$ & $341\pm51$ & $-0.9$ to 0.3 & &\\[0.5ex]
1399 total light & $-1.6\pm0.1$ & & & $-22.33\pm0.20$\footnote{including the
cD halo within $10\arcmin$, see Sect.~3.3} & $4.5 \times 10^{11}$\footnote{
derived from the $(M/L)_R = 3.83$ (van der Marel \cite{vdma}), $(V-R) = 0.58$
(Poulain \cite{poul88}), and $M_{V,\sun} = 4.76$}\\
1399 cD halo & $-1.0\pm0.2$ & 270 to 400\footnote{planetary nebulae
(Arnaboldi et al. \cite{arna})} & ? & $-21.65\pm0.20$ & $1.9 \times 10^{10}$ \\
1399 bulge & $-2.0\pm0.2$ & 200 (to 360) & $-0.7$ to $-0.1$ & $-21.50\pm0.20$ 
& $1.6 \times 10^{10}$ \\[0.5ex]
X-ray gas ($r < 0\fdg7$) & $-1.2\pm0.1$ & 390 to 460\footnote{derived from 
the temperature of the gas (Jones et al. \cite{jone97})} & $-0.6$ to 0.0 & &
$\simeq 2.7 \times 10^{11}$\footnote{Ikebe et al. (\cite{ikeb}), corrected
for a distance of 18.2 Mpc} \\
X-ray gas ($r < 10\arcmin$) & & & & & $\simeq 1.2 \times 10^{10 g}$ \\
\noalign{\smallskip}
\hline
\end{tabular}
\end{minipage}
\end{flushleft}
\end{table*}

Furthermore, it is interesting to note that also a large amount of
the intracluster gas (seen as X-ray halo) might have had its origin in
dwarf galaxies, which could have expelled their gas by supernova-driven
winds, or stripped off their gas (Trentham \cite{tren94}, Nath \&
Chiba \cite{nath}). In the Virgo cluster, for example, Okazaki et al.
(\cite{okaz93}) estimated that the amount of gas expelled from the E and S0
galaxies is not adequate to account for the total gas mass in the cluster.
Mac Low \& Ferrara (\cite{macl})
calculated that low mass dwarf galaxies can easily blow away metals from
supernovae which might enrich the halo gas.

\section{Properties of dwarf galaxies, GCs, and cD halo in the
Fornax cluster center}

The center of the Fornax cluster hosts the central
galaxy NGC~1399 with an extraordinarily rich globular cluster system and an
extended cD halo as well as a halo of X-ray emitting gas.
In the following we give a short review on the properties of the different
components that have to be considered in the picture of a common evolution.
In Table \ref{toverview} thoses properties are summarized: 
the slopes of the surface density profiles, the velocity
dispersion, and the ranges of metallicities. Furthermore, the absolute $V$
luminosities and estimated masses are given, if available.

\subsection{Dwarf galaxies in the Fornax cluster}

The most complete investigation of the Fornax dwarf galaxies was done by
Ferguson (\cite{ferg89}, Fornax Cluster Catalog (FCC)) as well as by Davies 
et al. (\cite{davie88}, and following papers: Irwin et al. \cite{irwi},
Evans et al. \cite{evan}). As we have shown in Paper~I the morphological
classification of Fornax members by Ferguson (\cite{ferg89}) is very reliable 
and nearly no dE has been missed within the survey limits as far as
we can judge from the comparison with our sample fields.
Thus, the following properties of the Fornax dwarf galaxies are 
mainly based on the FCC plus the additional new members as presented in Paper~I.

\begin{table*}
\caption{\label{tslopes} Power law slopes of the surface density profiles of 
Fornax galaxies taken from the FCC (Ferguson \cite{ferg89})}
\begin{flushleft}
\begin{tabular}{lccccc}
\hline\noalign{\smallskip}
 & all members & all dE/dS0 & nucleated & non-nucl. & E/S0 \\
\noalign{\smallskip}
\hline\noalign{\smallskip}
number & 292 & 192 & 97 & 95 & 27\\
$r < 0\fdg8$ & $-0.30\pm0.06$ & $-0.19\pm0.03$ & $-0.40\pm0.11$ & 
$-0.20\pm0.02$ & $-0.33\pm0.04$ \\
$0\fdg8 < r < 3\degr$ & $-1.51\pm0.13$ & $-1.52\pm0.28$ & $-2.04\pm0.17$ & 
$-1.06\pm0.58$ & $-1.49\pm0.56$ \\
total & $-1.04\pm0.23$ & $-1.03\pm0.26$ & $-1.20\pm0.25$ & $-0.86\pm0.45$ &
$-1.07\pm0.24$ \\
\noalign{\smallskip}
\hline
\end{tabular}
\end{flushleft}
\end{table*}

The spatial distribution of dEs in Fornax can be represented by a King profile
with a core radius of $0\fdg67\pm0\fdg1$ and a center located about $25\arcmin$ 
west of NGC~1399 (Ferguson \cite{ferg89}). In order to compare their surface
density profile with that of the GCS and the cD halo light we fitted power
laws to the radial distribution of the dEs and dS0s
in the extended FCC adopting NGC~1399 as the center.
For that we counted
galaxies brighter than $B_T = 19$ mag in 7 equi-distant rings from 0 to 
$3\degr$. We determined the slopes of the density profiles in the
inner ($r < 0\fdg8$) as well as in the outer ($0\fdg8 < r < 3\degr$) part.
The dividing radius of $0\fdg8$ is about the limit out to where the cD halo 
light and the gas envelope have been measured.
The results are summarized in Table \ref{tslopes}. In addition, we also give 
the mean slopes, when fitting a power law to the total profile, and the fitted
values for the giant galaxies. The nucleated dwarf galaxies have the steepest 
slope and
are more concentrated towards the central galaxy than the non-nucleated dE/dS0s.

\begin{table*}
\caption{\label{tfcclkf} Faint end slopes of the luminosity function for
different subsamples of the Fornax galaxy population}
\begin{flushleft}
\begin{minipage}{15.5cm}
\begin{tabular}{lrccc}
\hline\noalign{\smallskip}
 & $B_{\rm limit}$ & all members & dE/dS0 & non-nucleated \\
\noalign{\smallskip}
\hline\noalign{\smallskip}
$r < 3\fdg5$ & 19.5 & $-1.03\pm0.09$ & $-0.90\pm0.05$ & $-0.73\pm0.13$ \\
$r < 3\fdg5$ & 19.0 & $-1.23\pm0.08$ & $-0.81\pm0.07$ & $-0.51\pm0.18$ \\
$r < 0\fdg7$ & 19.5 & $-1.05\pm0.19$ & $-1.27\pm0.14$ & $-1.09\pm0.30$ \\
$r < 0\fdg7$ & 19.0 & $-1.16\pm0.25$ & $-1.20\pm0.17$ & $-2.01\pm0.28$ \\
$0\fdg7 < r < 2\fdg4$ & 19.5 & $-1.11\pm0.07$ & $-0.99\pm0.07$ & $-0.95\pm0.21$
\\ 
$0\fdg7 < r < 2\fdg4$ & 19.0 & $-1.19\pm0.09$ & $-0.92\pm0.10$ & $-0.63\pm0.76$ 
\\
F\&S88\footnote{Ferguson \& Sandage (\cite{ferg88})} ($r < 2\fdg4$) & 19.5 & 
$-1.32\pm0.09$ & $-1.08\pm0.19$ & - \\
\noalign{\smallskip}
\hline
\end{tabular}
\end{minipage}
\end{flushleft}
\end{table*}

The luminosity function (LF) of the Fornax dwarf galaxies was studied by 
Ferguson \& Sandage (\cite{ferg88}) in a region with radius smaller than 
$2\fdg4$, 
centered on NGC~1399. They found that the nucleated dwarf ellipticals (dE,Ns) 
as well as the dwarf lenticular (dS0) galaxies are 
brighter than the non-nucleated dEs. Further, the faint end slope of the
dE/dS0 LF, fitted by a Schechter (\cite{sche}) function, is quite flat
($\alpha = -1.08\pm0.10$) compared to other clusters like Virgo ($\alpha = 
-1.31\pm0.05$) or Centaurus ($\alpha = -1.68\pm0.56$, Jerjen \& Tammann 
\cite{jerj97c}).
Table \ref{tfcclkf} summarizes the results for the faint end slopes of
Schechter function fits to different subsamples of the extended FCC.

Colors and metallicities of dwarf galaxies in Fornax have been studied by
photometric as well as by spectroscopic means (e.g. Caldwell \& Bothun 
\cite{cald87b}, Bothun et al. \cite{both}).
Spectroscopically determined metallicities seem to be
consistent with the picture that the bluer dwarfs are the more metal-poor ones.
The metallicity range for 10 bright dE,Ns is $-1.5 < [Fe/H] < -0.8$ dex (Held
\& Mould \cite{held}). The metallicities derived from Washington photometric
indices for 15 LSB dwarfs are of the same order (Cellone et al. \cite{cell94}).
Concerning ages, all investigated dwarfs possess an old stellar population,
some of them a contribution of intermediate-age stars, and only few have signs
of recent or ongoing star formation (Held \& Mould \cite{held}, Cellone \& 
Forte \cite{cell96}).
It seems that the Fornax dEs share the same characteristics as the Local Group 
dSph population (e.g. review by Grebel \cite{greb97}).

Radial velocity measurements of 43 Fornax dwarfs ($18 > B_t > 15$ mag) by 
Drinkwater et al. (\cite{drin97})
result in a velocity dispersion of $\sigma_v = 490$ km~s$^{-1}$, significantly
larger than that of 62 giants ($B_t < 15$ mag), $\sigma_v = 310$ km~s$^{-1}$.
According to the authors, this difference cannot be explained by 
measurement errors.

\subsection{The central globular cluster system}

The globular cluster system of NGC 1399 is one of the best investigated GCSs
outside the Local Group.
The total number of GCs is about $N_{\rm GC} = 5800\pm500$ (Kissler-Patig et 
al. \cite{kiss97a}, Grillmair et al. \cite{gril98}) within a radius of
$10\arcmin$ from the galaxy center. This is about 10 times the number
of GCs in the other Fornax ellipticals, $300\pm60 < N_{tot} < 700\pm100$.
Adopting a distance of 
18.2 Mpc or $(m-M)_0 = 31.3$ mag to NGC 1399 (Kohle et al. \cite{kohl},
recalibrated with new distances of Galactic GCs, Gratton et al. \cite{grat})
the absolute magnitude of NGC~1399 is $M_V = -21.75$ mag when taking the
apparent magnitude values from the literature (Faber et al. \cite{fabe89},
RC3: de Vaucouleurs et al. \cite{devo91}). This corresponds to a specific 
frequency of $S_N = 11.6\pm2.0$. If the light of the cD halo within $10\arcmin$
is taken into account (see Sect.~3.3), $S_N$ is reduced to $6.8\pm2.0$ 
($M_{V,{\rm tot}} = -22.33$ mag). However, distinguishing a cD halo and a bulge 
component in the galaxy light, $S_N$ for the cD halo would be about $10\pm1$
(assuming $S_N=3.2$ for the bulge, see Sect.~9), the average value of the other 
early-type Fornax galaxies (Kissler-Patig et al. \cite{kiss97a}).
Thus, the building up of the GCS of the cD halo component must have been 
very efficient.

The color distribution of the GCs around NGC 1399 is very broad compared
to most other GCSs in Fornax ellipticals and can only be explained by a
multimodal or perhaps just a bimodal GC population (e.g. Ostrov et al. 
\cite{ostro}, Kissler-Patig et al. \cite{kiss97a}, and Forbes et al.
\cite{forb97}). Spectroscopic analysis of 18 GCs by Kissler-Patig et al.
(\cite{kiss98a}) shows a metallicity range between $-1.6$ 
and $-0.3$ dex (with possible peaks at $-1.3$ and $-0.6$ dex), and two
exceptional GCs at about 0.2 dex, located in the red (metal rich)
tail of the color distributions. The comparison of the line indices with
theoretical evolutionary models suggests that most of the GCs are older than
at least 8 Gyr. If one fits the GC color distribution with two Gaussians, 
the number ratio of metal rich (red) to metal poor (blue) GCs is about 1:1 
(Forbes et al. \cite{forb97}).

The radial extension of the GCS around NGC 1399 can be traced out to about
$10\arcmin$ ($\simeq 53$ kpc). The slope of the GC surface density profile,
$\rho \propto r^\alpha$, is about $\alpha = -1.5\pm0.2$, when taking the
average of the published values. Forbes et al. (\cite{forb97}) found that the
distribution of the blue GC subpopulation is even flatter ($\alpha \simeq 
-1.0\pm0.2$), whereas the red GCs are more centrally concentrated ($\alpha 
\simeq -1.7\pm0.2$), comparable to the slope of the galaxy light ($\alpha =
-1.6\pm0.1$). See Fig.~\ref{fsdslope} for a schematic overview.

Radial velocities of 74 GCs around NGC~1399 have been measured (Kissler-Patig
et al. \cite{kiss99a}, Minniti et al. \cite{minn98}, Kissler-Patig et al. 
\cite{kiss98a}). The velocity 
dispersion for the whole sample is $\sigma_v = 373\pm35$ km~s$^{-1}$. 
No differences can be seen between the red and blue subpopulations.
However, there exists a radial dependence of the velocity dispersion in the
sense that $\sigma_v$ rises from $263\pm92$ to $408\pm107$ km~s$^{-1}$
between $2\arcmin$ and $8\arcmin$ (Kissler-Patig et al. \cite{kiss99a}).

\subsection{cD halo and bulge}

The galaxy light of NGC~1399 follows an extended cD profile (Schombert 
\cite{scho86}, Killeen \& Bicknell \cite{kill}) out to a radial distance
of about 34 arcmin from the galaxy center ($\Sigma_B = 28$ mag isophotal
surface brightness level). This is about 180 kpc in Fornax distance (18.2 Mpc)
and comparable to the extent of the X-ray envelope (Ikebe et al. \cite{ikeb},
Jones et al. \cite{jone97}).

\begin{figure}
\psfig{figure=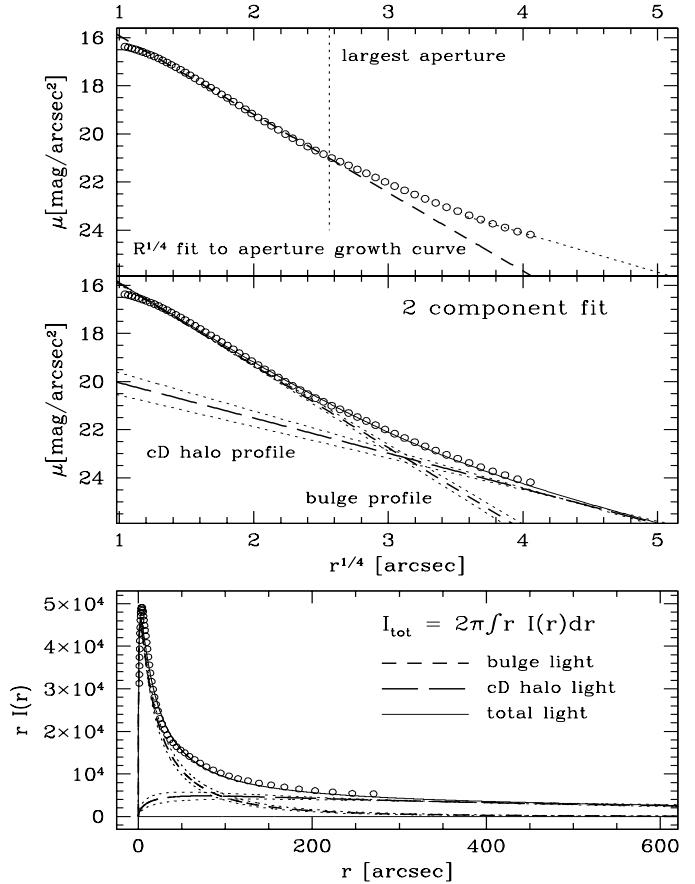,height=11.8cm,width=8.6cm
,bbllx=13mm,bblly=61mm,bburx=134mm,bbury=249mm}
\caption{\label{fcdprof} The upper two panels show the surface brightness 
profile (open circles) of NGC~1399 plotted versus $r^{1/4}$.
The change of the slope to a flat cD halo is clearly visible.
In the uppermost panel, the dashed line corresponds to the $R^{1/4}$ law
extrapolation of an aperture growth curve used to derive the apparent
magnitude $V = 9.55$ mag (Faber et al. \cite{fabe89}, RC3). The largest
aperture given by Burstein et al. (\cite{burs84b}) is indicated. The
dotted extension of the surface brightness profile at the faint end 
represents the slope between $2\arcmin$ and $10\arcmin$ by Killeen \& Bicknell
(\cite{kill}).
The dashed lines in the second panel represent the single components of
the cD halo and the underlying bulge light, when fitting the sum of two de
Vaucouleur laws (solid line). The dotted lines give the ranges for possible 
fits.
In the lowermost panel, $r~I(r)$ is plotted versus the radial distance to the
center of NGC~1399. Again, the dashed curves describe the single components 
and the dotted ones the possible ranges. Within $10\arcmin$ the total
dotted ones the possible ranges. Within $10\arcmin$ the total luminosity
luminosity $I_{\rm tot}$ of the cD halo light is slightly higher than that of 
the bulge light.
}
\end{figure}

The determination of the stellar population parameters of the outer cD
halo, like accurate photometric colors, metallicity or velocity dispersion,
is very difficult due to the low surface brightness. Long slit spectra have
been taken for the stellar bulge population within a radius of about $1\farcm5$
from the center of NGC 1399 (Franx et al. \cite{fran}, Bicknell et al.
\cite{bick}).
The velocity dispersion is about 200 km~s$^{-1}$ at $1\farcm5$ and rises within 
the central
$10\arcsec$ to a central value of about 360 km~s$^{-1}$. Besides the GCS, an 
useful tracer for
the stellar population at larger radii is the population of
planetary nebulae (PNe). Arnaboldi et al. (\cite{arna}) studied the kinematics
of 37
PNe out to a radius of $4\farcm5$. They found an increase in the velocity
dispersion with increasing radius from 269 km~s$^{-1}$ for $r < 2\farcm6$ to
405 km~s$^{-1}$ for $2\farcm6 < r < 4\farcm5$ (18 of the 37 PNe).

\subsubsection{Luminosity and surface brightness profile}

\begin{figure}
\psfig{figure=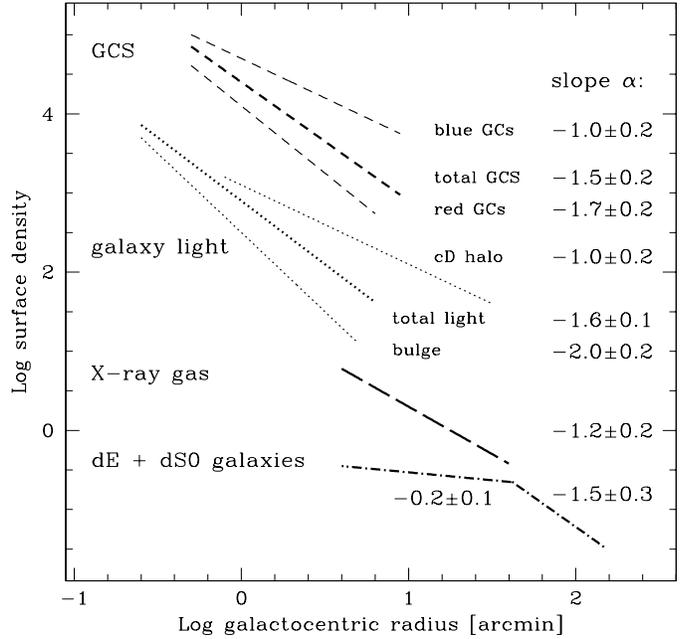,height=8.6cm,width=8.6cm
,bbllx=15mm,bblly=56mm,bburx=195mm,bbury=246mm}
\caption{\label{fsdslope} Schematic overview of the surface density or
brightness profiles of different subpopulations in the Fornax cluster and
their extensions. The profiles are arbitrarily shifted in the ordinate axis.
Also the units are arbitrarily chosen.
The profiles of the GCs and galaxies are surface density profiles, the one
of the galaxy light is a surface brightness profile, and the one of the X-ray
gas is again a particle (number) density profile (see text for more details).
The slopes $\alpha$ ($\rho \propto r^\alpha$) and their uncertainties are
indicated on the right hand side.
}
\end{figure}

In this subsection we divide the light profile of NGC~1399 into a cD halo and
a bulge component in order to compare their characteristics with those of the
GCS and the dwarf galaxy population.
We determined the absolute luminosity of the cD halo in the following way: 
in the $\mu$--$r^{1/4}$ plot (Fig.~\ref{fcdprof}, upper panels) one can see 
that the SB profile of NGC 1399 (determined from the NE CCD field F2) changes 
its slope at about $50\arcsec$. We fitted the total profile by the sum of 
two de Vaucouleurs laws:\\
$\mu (r) = ZP_{\rm cal} - 2.5\cdot log [I^0_{\rm gal}\cdot 
exp(-(r/\alpha_{\rm gal})^{1/4}) + I^0_{\rm cD}\cdot
exp(-(r/\alpha_{\rm cD})^{1/4})]$\\
The steeper, more concentrated profile represents the luminosity of the 
bulge without cD halo, whereas the flatter, more extended profile contains
the light of the cD halo. The total luminosity of each component is
$I^{\rm tot}_{\rm gal,cD} = 2\pi \int r I_{\rm gal,cD}(r)dr$. We restricted our
calculations to within a radius of $10\arcmin$ where the number of detected 
GCs fades
into the background. The dashed lines in Fig.~\ref{fcdprof} represent the
``best'' fit (data points inside $1\farcs5$ radius have been omitted). The
dotted lines give the ranges for possible fits. The surface density slopes
of both profiles are $\alpha = -2.0\pm0.2$ and $\alpha = -1.0\pm0.2$ 
respectively, if $\Sigma(r) \propto r^\alpha$. The slope of the combined
profile is $\alpha = -1.6\pm0.1$ (see also Fig.~\ref{fsdslope}). 

In the literature one finds an apparent magnitude for NGC~1399 of
$V = 9.55$ mag (Faber et al. \cite{fabe89}, RC3: de Vaucouleurs et al. 
\cite{devo91}, adopting a mean $(B - V)$ color of 1.0 mag, Goudfrooij et al.
\cite{goud94b}). This magnitude is derived from an aperture growth curve
extrapolation. with a maximum aperture of diameter $1\farcm5$
(Burstein et al. \cite{burs84b}).
A 1-component fit of an $R^{1/4}$ law within $1\farcm5$ (the largest aperture in
Burstein et al. \cite{burs84b}) is shown in Fig.~\ref{fcdprof}
(uppermost panel). Adopting an absolute magnitude of $M_V = -21.75$ mag for the
integrated light under this profile, the total luminosity of the
bulge light from the 2-component fit (middle panel) is $M_{V,{\rm bulge}} = 
-21.50\pm0.20$ mag (about 80\% of the luminosity given in the literature).
The luminosity for the
cD halo is $M_{V,{\rm cD}} = -21.65\pm0.2$ mag and for the whole system
within $10\arcmin$ $M_{V,{\rm tot}} = -22.33\pm0.2$ mag. 

Another
check for the correct proportion of the luminosities of the different 
components can be made
by comparing the integrated flux within an aperture of $1\farcm5$ with the
total flux within $10\arcmin$. Adopting $V = 10.30$ mag for the $1\farcm5$
aperture (Burstein et al. \cite{burs84b}), we derive $M_{V,{\rm tot}} = 
-22.27\pm0.2$ mag for the whole system, in excellent agreement with the value 
given above.

Note that the total luminosity of the cD halo is about 180 times the luminosity 
of a typical dwarf galaxy with $M_V = -16.0$ mag or 2.2 times the total 
luminosity of the present dEs and dS0s in Fornax.

\section{Comparison of corresponding properties}

In the previous section we presented the properties of the different components
of NGC~1399 and the galaxy population in the center of the cluster. 
In Figure~\ref{fsdslope} we give a schematic overview of the surface 
density/brightness (SD/SB) profiles of the different components and 
their extension. The profiles are arbitrarily shifted in the ordinate axis.
The surface densities of GCs and
galaxies are number densities, whereas the profile of the galaxy light is a 
surface brightness profile. Nevertheless, this is comparable to the number 
density profile, if one assumes similar stellar populations and, accordingly,
similar $M/L_V$ ratios. The surface density of the X-ray gas is again a particle
(number) density, $n_{\rm gas} \propto r^{\gamma}$. The gas density profile 
is derived from the surface brightness distribution of the X-ray gas $S(r) 
\propto r^{\tau}$ under the assumption of isothermal conditions ($\tau = 
(\gamma+1)/2$) for a radius larger than about 10 kpc (Jones et al. 
\cite{jone97}).

The plot shows that the profile slopes of the blue GC population and the cD halo
light are strikingly similar, whereas the distribution
of dE/dS0 galaxies is somewhat flatter in the central $0\fdg7$ and slightly
steeper outside. 
Interestingly, the surface density profile of the X-ray gas is also very similar
in slope and extension to the cD halo light and the blue GCs.
In contrast, the profile of the bulge light of NGC~1399 is significantly
steeper than all other profiles.

The same behaviour can be seen in the velocity dispersion.
It is comparable for dwarf galaxies and GCS, whereas the stars
in the stellar bulge have a lower $\sigma_v$ (see also Minniti et al. 
\cite{minn98}, Kissler-Patig et al. \cite{kiss99a}).
This agreement in morphological and dynamical properties of GCs, the cD halo 
light, and perhaps also the gas particles might suggest that 
these components share a common history (or origin).
In the next section we describe some scenarios that might have happened
when dwarf galaxies interacted with the central cluster galaxy.

\section{Disruption, accretion and stripping of dwarf galaxies}

What are the possible consequences, when dwarf galaxies of different types
interact with the central galaxy, especially with respect to
the formation of a cD halo and a rich GCS?

We make a distinction between two main cases:\\
(1) the infall of gas-poor dwarfs, for example dwarf ellipticals, where
only the existing stellar component is involved in the interaction process, and
\\
(2) the infall of gas-rich dwarfs, where the interaction of the gas
has to be considered and might play an important role in the formation of
new stellar populations

A further sub-division of these cases is:\\
(a) the dwarf galaxy will be totally dissolved in the interaction process\\
(b) only parts of the dwarf galaxy (for example gas and/or globular clusters)
will be stripped during the passage through the central cluster region\\
(c) the dwarf galaxy neither loses gas nor stars nor clusters to the cluster
center, but might change its morphological shape because of tidal interactions
(for example getting more compact or splitting into two).

In the next subsections we discuss the possible consequences for the different 
cases.

\subsection{Gas-poor dwarfs}

\noindent
(1a): in this case the stellar population of the dwarf galaxy will be
disrupted in tidal tails and the stellar light will be smeared out 
in the potential well of the cluster center. Most affected by this process
are the faintest dEs (or dSphs, Thompson \& Gregory \cite{thom}).
In clusters with a low velocity dispersion or at the bottom of a local
potential well in a rich cluster (Zabludoff et al. \cite{zabl}), the light of 
several dissolved dwarfs may form an extended, diffuse cD halo. Existing GCs
of the dwarfs will survive and contribute to the central GCS.
In the Local Group, an example for this scenario may be the Sagittarius
dSph which is dissolving into our Galaxy, adding 4 new GCs to the GCS of the 
Milky Way
(Da Costa \& Armandroff \cite{daco95}). However, only few dwarf galaxies 
with a very rich GCS compared to their luminosity are known
(Miller et al. \cite{mill}, Durrell et al. \cite{durr}).
In Sect.~6 we estimate under which
conditions the accretion of gas-poor dwarfs and their GCS can increase $S_N$
of a central GCS. 

Finally, the nuclei of dE,Ns
can survive the dissolution of their parent galaxy and
may appear as GCs (Zinnecker et al. \cite{zinn}, Bassino et al \cite{bass}). 
The nuclear magnitudes of all Virgo dE,Ns (Binggeli \& Cameron \cite{bing91}),
for example, fall in the magnitude -- surface brightness sequence is defined
by the GCs (e.g. Binggeli \cite{bing94}).\\

\noindent
(1b): like in the case 1a the stripped stars and GCs will be distributed around
the central galaxy.
In this case the question arises on how large the number of stripped GCs
is compared to the luminosity of the stripped stellar light. If GCs could be 
stripped from regions with a high local $S_N$, this would also increase $S_N$
of the central GCS. According to model calculations by Muzzio et al. 
(\cite{muzz84}, see also review by Muzzio \cite{muzz87b}), the tidal accretion 
of GCs and stars can be an important process in the dynamical evolution of
GCSs in galaxy clusters. 

In some galaxies the GCS is more extended than the underlying
stellar light, which has the consequence that the local $S_N$ increases with
galactocentric distance; for example NGC 4472 has a global $S_N$ of 5.5
and a local $S_N$ larger than 30 at 90 kpc (McLaughlin et al. \cite{mcla94b}). 
Forbes et al. (\cite{forb97}) and Kissler-Patig et al. (\cite{kiss99a})
suggest that the stripping of the outermost GCs and stars from 
such a galaxy naturally increases the $S_N$ of the central GCS.
It would be interesting to investigate whether this is also true for the 
GCSs of dwarf galaxies.

Furthermore, it would be interesting to know how the
tidal stripping process changes the shape of the remaining galaxy. Kroupa
(\cite{krou}) simulated the interaction of a spherical low-mass galaxy with a
massive galactic halo and found that the model remnants share the properties
of dwarf spheroidals. On the other hand, M32 may be
an example for a tidally compressed remnant, whose GCs have been stripped
(e.g. Faber \cite{fabe73}, Cepa \& Beckman \cite{cepa}).\\

\begin{table*}
\caption{\label{tmcgc} Initial conditions for GCSs of dwarf galaxies in the
Monte Carlo simulations. $N_{\rm gc}$ is the possible number of GCs in the
magnitude bin. $S_N$ gives the range of specific frequencies that can be
achieved with the number of GCs
}
\begin{flushleft}
\begin{tabular}{lcccccccc}
\hline\noalign{\smallskip}
bin & $< -15.5$ & $-15.5:-14.5$ & $-14.5:-13.5$ & $-13.5:-12.5$ & $-12.5:-11.5$
&
$-11.5:-10.5$ & $-10.5:-9.5$ & $-9.5:-8.5$\\
\noalign{\smallskip}
\hline\noalign{\smallskip}
$N_{\rm gc}$ & & 6--20 & 2--9 & 1--5 & 0--2 & 0--1 & 0--1 & 0--1 \\
$S_N$ & 4.5 & 1.8--16.6 & 1.7--18.8 & 2.1--28.8 & 0--29 & 0--36 & 0--91
& 0--251 \\
\noalign{\smallskip}
\hline
\end{tabular}
\end{flushleft}
\end{table*}
 
\noindent
(1c): in this case the dwarf galaxy does not contribute to the formation of cD
halo and central GCS. However, as in 1b one might speculate about the change
of the morphological shape after a passage of the galaxy through the cluster
center.\\

Note that, except in their nuclei, the metallicity of GCs in dEs as
well as the metallicity of the bulk of their stars
is very low ($-2.5 < [Fe/H] < -1.0$ dex, see the review on Local Group
dwarfs by Hodge \cite{hodg94b}).
Therefore, stripped GCs from these galaxies will only contribute to
the metal-poor population of the central GCS. And accordingly, the cD 
halo should have quite a blue color.

\subsection{Gas-rich dwarfs}

\noindent
(2a): for the stellar population and GCs see 1a. The infalling gas will
experience the thermal pressure of the hot medium in the central galaxy.
The densities can be high enough that star formation occurs and the formation
of many dense and compact star clusters is possible (e.g. Ferland et al.
\cite{ferl}). As mentioned in Sect.~3.4, stripped gas that was not
converted into
stars may contribute to the
intracluster X-ray gas in the cluster center (see Nath \& Chiba \cite{nath}).
The open questions here are, how many ``new'' star clusters will survive the
further cluster center evolution, and how large the number of surviving
clusters is compared to the light of newly formed stars which contribute
to the total light of the central galaxy and/or cD halo. In other words, it is
unclear whether the formation of new GCs is so efficient that it can increase
$S_N$ of the central GCS.

Some constraints/estimations that can be made from observations of very young
star clusters in merging galaxies and starburst galaxies are presented in
Sect.~7.\\

\noindent
(2b): the stripping of a gas-rich galaxy mainly affects the gas component
that may form new stars and clusters as mentioned in case 2a. Nulsen
(\cite{nuls}, see also Ferguson \& Binggeli \cite{ferg94}) estimated a typical
mass loss rate from infalling dwarfs of\\
\centerline{$\dot{M} = 7.4\cdot10^{-2}M_{\sun} {\rm yr}^{-1} n r_{\rm kpc}^2
\sigma_{\rm km~s^{-1}}$,}
where $n$ is the gas density in the cluster, $r_{\rm kpc}$ the dwarf galaxy
radius in kpc, and
$\sigma_{\rm km~s^{-1}}$ the velocity dispersion of the cluster.
A stripping time scale for a typical dwarf irregular (dI), $r = 4$ kpc,
gas mass
$M_{\rm gas} = 10^8 M_{\sun}$, and $\sigma = 400$ km~s$^{-1}$ (Fornax) is
$t_S = 0.5$ Gyr, when adopting $n = 8.2\cdot10^8$ cm$^{-3}$ for the central
gas density of the Fornax cluster (Ikebe et al. \cite{ikeb}).
The fact that dEs are more concentrated towards cluster
centers than dIs is interpreted as a result of this stripping
scenario (dEs being the remnants of stripped dIs, e.g. Lin \& Faber \cite{lin},
Kormendy \cite{korm}). Furthermore, non-nucleated dEs have a quite low GC $S_N$
in contrast to dE,Ns, for which $S_N$ increases with decreasing luminosity
(Miller et al. \cite{mill}).
One might speculate that not only gas but also GCs have been stripped,
whereas the dE,Ns have a different evolution history.\\

\noindent
(2c): the passage of a gas-rich dwarf through the intergalactic gas in a
cluster might trigger star formation in the dwarf galaxy itself (Silk et al.
\cite{silk}) and might enrich its
GCS (see Sect.~7). Ferguson \& Binggeli (\cite{ferg94})
suggest that a galaxy falling into the cluster for the first time in the
present epoch encounters such high densities that stars can form
long before ram pressure becomes efficient. A further close
passage then can result in the cases 2a or 2b.\\

In all theses cases there are no restrictions for the metallicity of newly 
formed
GCs and stars. Their metallicity depends on the gas enrichment history of the
accreted or stripped galaxies themselves. Of course, a lower limit is the
metallicity of the interstellar matter of the dwarf galaxy.
As an estimation for this lower limit can serve the metallicity of the
young populations in the Local Group dSphs and irregulars.
It seems that the secondary stellar population has metallicities more metal-rich
than $[Fe/H] = -1.2$ dex, in some cases even up to solar values
(e.g. Grebel \cite{greb97}).

\section{Enhancement of $S_N$ by accretion of gas-free dwarf galaxies}

The possibility that the accretion of gas-poor dwarfs can increase $S_N$
of the central GCS requires that the $S_N$ value of a large number of accreted
dwarfs themselves is very high.
Only few examples of dwarf galaxies with very high GC frequencies are known.
In the Local Group the Fornax and Sagittarius dwarf spheroidals have
extraordinarily high $S_N$ values: $29\pm6$ and $25\pm9$ respectively
(Durrell et al. \cite{durr}).
Their absolute luminosities are about $M_V = -13$ mag.
Durrell et al. (\cite{durr}) found in the Virgo cluster two dE,Ns
fainter than $M_V = -15.5$ mag that have GC specific frequencies
in the order of $S_N = 14\pm8$.
Recently Miller et al. (\cite{mill}) found that
exclusively dE,Ns can possess high $S_N$ GCSs, whereas dEs have
``normal'' values. In this respect, it is worthwhile noting that the 
nuclei of dE,Ns could be merged globular clusters, and thus the $S_N$ 
of these galaxies might have been even higher in the past.
It seems that all of the high $S_N$ dwarf galaxies belong to the faint
luminosity end of the dwarf galaxy population.
Thus, their total numbers of GCs are very small, $4 < N_{tot} < 20$ and it
might reflect the stochastic effect, where a low mass dwarf is able to
produce no, 1, 2, or several clusters.

\begin{figure}
\psfig{figure=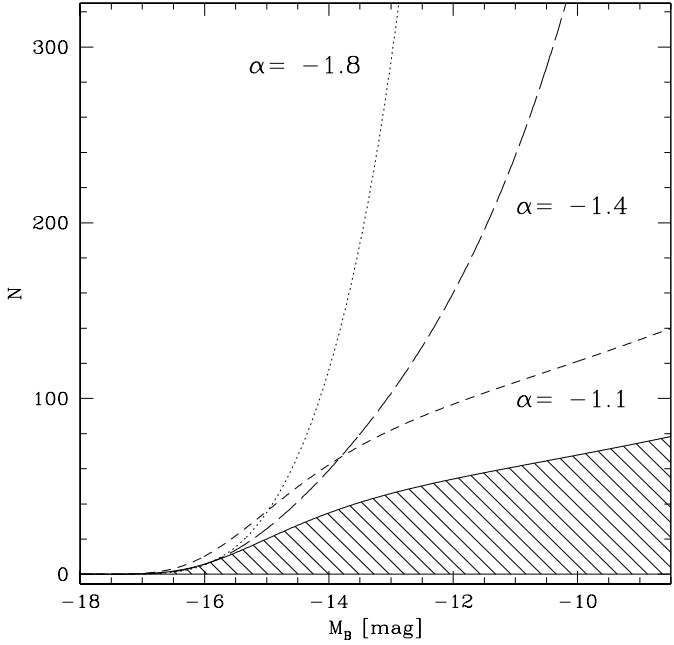,height=8.6cm,width=8.6cm
,bbllx=9mm,bblly=57mm,bburx=195mm,bbury=246mm}
\caption{\label{fmclkf} The dotted and dashed lines are initial Schechter-type
luminosity functions with different slopes as indicated. The hatched area
represents the present day luminosity function for dE and dS0 galaxies in
Fornax. Note that for $M_B > -12.0$ mag the galaxy counts in Fornax are
incomplete.}
\end{figure}

We tested with Monte Carlo (MC) simulations the possibility that 2500 GCs
were captured by gas-poor dwarf galaxy accretion in the center of the
Fornax cluster. The number 2500 comprises about the blue GC subpopulation.
We assumed that galaxies with absolute luminosities in the range $-18.0 <
M_B < -8.5$ mag have been accreted. Each galaxy contains GCs according to 
its luminosity. For galaxies with $-18.0 < M_B < -15.5$ we adopted a mean
GC specific frequency of $S_N = 4.5$ (Durrell et al. \cite{durr}).
In the fainter magnitude bins, the number of GCs was chosen randomly in such
a way that the ranges of observed $S_N$ (Miller et al. \cite{mill}) were 
reproduced.
In Table \ref{tmcgc} the initial conditions for the dwarf GCSs are summarized.
We simulated 3 cases, a very optimistic one (simulation run~1, very faint
dwarfs can also possess GCs), a pessimistic case (run~3), where no dwarf 
fainter than
$M_B = -12.5$ can possess GCs (as it seems to be the case for the Local Group
dSphs), and a medium case (run~2, dwarfs fainter then $M_B = -10.5$ can not
possess any GC).
However, if faint dEs are already stripped dwarf galaxies,
the simulation run~1 or 2 seem more reasonable.

\begin{table}
\caption{\label{tmcsim} Results of MC simulations for the disruption of
gas-poor dwarfs in the center of the Fornax cluster. In simulation run~1
all dwarfs down to $M_B = -8.5$ mag can possess GCs (see also
Table \ref{tmcgc}), whereas in run~2 and 3 only dwarfs brighter than $M_B =
-10.5$ and $-12.5$ mag can possess GCs, respectively. $\alpha$ and $M^*$ are
the parameters for the initial LF, $N_{\rm tot}$ is the number of galaxies
that has been disrupted to account for 2500 GCs, $S_{\rm N,dw}$ is the
specific frequency of captured GCs compared to the disrupted galaxy light,
$\Delta L = L_{\rm dw}/L_{\rm cD}$ is the ratio of disrupted galaxy light 
$M_{\rm B,t}$ and 
the present day cD halo light, and $\overline{[Fe/H]}$ is the mean metallicity 
of the accreted GCs}
\begin{flushleft}
\begin{tabular}{crr@{\hspace{4mm}}rrrrr}
\hline\noalign{\smallskip}
Run & $\alpha$ & $M^*$ & $N_{\rm tot}$ & $M_{\rm B,t}$ & $S_{\rm N,dw}$ &
$\Delta L$ & $\overline{[Fe/H]}$ \\
\noalign{\smallskip}
\hline\noalign{\smallskip}
1 & -1.1 & -15.3 &  1000 & -20.6 &  7.1 & 1.0 & -1.7 \\
  & -1.1 & -16.3 &   450 & -20.9 &  5.5 & 1.3 & -1.5 \\
  & -1.1 & -17.3 &   260 & -20.9 &  5.0 & 1.3 & -1.3 \\
  & -1.4 & -15.3 &  3180 & -20.0 & 15.0 & 0.5 & -2.0 \\
  & -1.4 & -16.3 &  1630 & -20.6 &  7.7 & 1.0 & -1.7 \\
  & -1.4 & -17.3 &   950 & -20.8 &  5.9 & 1.1 & -1.5 \\
  & -1.8 & -15.3 &  4050 & -19.6 & 22.3 & 0.4 & -2.1 \\
  & -1.8 & -16.3 &  3490 & -19.9 & 15.7 & 0.5 & -2.0 \\
  & -1.8 & -17.3 &  3060 & -20.1 & 12.4 & 0.6 & -1.9 \\
2 & -1.1 & -15.3 &  1190 & -20.7 &  6.7 & 1.0 & -1.7 \\
  & -1.1 & -16.3 &   513 & -20.9 &  5.3 & 1.3 & -1.5 \\
  & -1.1 & -17.3 &   248 & -20.9 &  4.9 & 1.3 & -1.3 \\
  & -1.4 & -15.3 &  6330 & -20.5 &  9.5 & 0.9 & -1.9 \\
  & -1.4 & -16.3 &  2340 & -20.7 &  6.4 & 1.0 & -1.6 \\
  & -1.4 & -17.3 &  1240 & -20.8 &  5.6 & 1.1 & -1.5 \\
  & -1.8 & -15.3 & 13770 & -20.5 &  9.2 & 0.9 & -1.9 \\
  & -1.8 & -16.3 &  9670 & -20.6 &  7.9 & 1.0 & -1.8 \\
  & -1.8 & -17.3 &  7530 & -20.7 &  7.1 & 1.0 & -1.7 \\
3 & -1.1 & -15.3 &  1200 & -20.8 &  5.9 & 1.1 & -1.6 \\
  & -1.1 & -16.3 &   480 & -20.9 &  5.1 & 1.3 & -1.5 \\
  & -1.1 & -17.3 &   270 & -21.0 &  4.8 & 1.4 & -1.3 \\
  & -1.4 & -15.3 &  9020 & -21.1 &  5.2 & 1.5 & -1.8 \\
  & -1.4 & -16.3 &  2380 & -21.0 &  5.2 & 1.4 & -1.6 \\
  & -1.4 & -17.3 &  1150 & -21.0 &  4.9 & 1.4 & -1.4 \\
  & -1.8 & -15.3 & 23810 & -21.5 &  3.8 & 2.2 & -1.8 \\
  & -1.8 & -16.3 & 11740 & -21.2 &  4.6 & 1.7 & -1.7 \\
  & -1.8 & -17.3 &  8250 & -21.1 &  4.7 & 1.5 & -1.6 \\
\noalign{\smallskip}
\hline
\end{tabular}
\end{flushleft}
\end{table}

We started our simulations with an initial Schechter-type LF with
a given characteristic luminosity $M^*$ and faint end slope $\alpha$.
Then we ``disrupted'' galaxies of randomly chosen luminosities as long as
2500 GCs have been accumulated, considering the condition that the final
LF resembles the present one of the Fornax cluster. 
We have chosen the following
initial faint end slopes: $\alpha = -1.1$ (the present day faint end slope of
dE/dS0s), $\alpha = -1.4$, and $\alpha = -1.8$. We varied the characteristic
luminosity between $M^* = -15.3$ (present day), $M^* = -16.3$, and
$M^* = -17.3$. The brighter $M^*$ the higher is the fraction of disrupted
dwarf galaxies at the bright end of the LF.
Figure \ref{fmclkf} shows the initial LFs with different
slopes compared to the present day LF (hatched area).
For each simulation we calculated the total number of disrupted dwarfs
$N_{\rm tot}$, their total luminosity $M_{\rm B,tot}$, the fraction of their
light compared to the cD halo light $\Delta L = L_{\rm dw}/L_{\rm cD}$, 
and the specific frequency of GCs compared to the disrupted stellar light
$S_{\rm N,dw}$.
In addition, we estimated the mean metallicity of the accreted GCs.
For galaxies brighter than $M_V = -13$ mag, we adopted the 
metallicity--luminosity relation given by C\^ot\'e et al. (\cite{cote}),
$\overline{[Fe/H]} = 2.31 + 0.638\cdot M_V + 0.0247\cdot M_V^2$. For
the fainter dwarfs, a metallicity--luminosity relation was derived from
a linear regression to the C\^ot\'e et al. data with $M_V < -13$ mag, 
$\overline{[Fe/H]} = -0.10\cdot M_V -3.13$. With this relation, the GCs of
the faintest dwarfs in our simulations, $M_B = -8.5$ mag, have a mean
metallicity of about $\overline{[Fe/H]} = -2.3$ dex.
Table \ref{tmcsim} summarizes the results of our simulations.

The MC simulations show that high $S_N$ values around 10
can only be achieved under the assumption that dwarf galaxies fainter
than $M_B = -10.5$ mag 
can possess at least one GC and that the faint end slope
of the initial LF is at least as steep as $\alpha = -1.4$. 
The dissolved light then comprises
about 60-90\% of the present day cD halo light (within $10\arcmin$).
However, the number of dissolved galaxies in theses cases,
3000--14000, is very high.
It has to be shown whether theoretical simulations
of cluster evolution can reproduce such high destruction rates, when
including also very low mass dwarf galaxies.
Moreover, the mean metallicity of the accreted GCs for the cases with
high $S_N$ values would be about 0.5 dex lower than the observed metal-poor
peak at $-$1.3 dex.
In Sect.~9 we discuss, whether the mixture of the presented accretion
process with stripping of GCs and new formation from infalling gas can
explain the luminosity of the cD halo together with the observed $S_N$ of
the central GCS in the Fornax cluster.

\section{Efficiency of new cluster formation by the accretion of gas-rich
dwarf galaxies}

In this section we assume that the gas of stripped dwarfs forms stars and 
clusters with the same proportion as has been determined for merging
and starburst galaxies.
Many examples for young GC candidates in mergers are known: e.g.
NGC 3597 (Lutz \cite{lutz}), NGC 1275 (Holtzman et al. \cite{holt}, 
N{\o}rgaard-Nielsen et al. \cite{norg}, NGC 5018 (Hilker \& Kissler-Patig
\cite{hilk96}), NGC 7252 (Whitmore et al. \cite{whitm93}, Schweizer \&
Seitzer \cite{schw93}, Whitmore \& Schweizer \cite{whitm95a}).
The number of newly formed clusters differs from case to 
case and seems to depend on the amount of gas that is involved in the merging
process. The precondition that is needed to form a bound cluster is a cold
gas cloud with high density in its core (e.g. Larson \cite{lars}). 
Furthermore, the 
local star formation efficiency has to be very high and has to occur on a 
short timescale in order to avoid an early disruption by strong stellar 
winds and by supernova explosions of the most massive stars (Brown et al. 
\cite{brow}, Fritze-v.~Alvensleben \& Burkert \cite{frit95}).
The best candidates for the progenitors of the clusters are the massive, 
embedded cores of (super) giant molecular clouds (e.g. Ashman \& Zepf 
\cite{ashm92}, Harris \& Pudritz \cite{harr94b}).
Elmegreen et al. (\cite{elme}) have shown that large
molecular cloud complexes can form in interacting systems. The high densities
in the cores that are necessary for the cluster collapse can be induced by 
direct cloud-cloud collision as well as by an increase of the ambient gas
pressure as a result of a merger (Jog \& Solomon \cite{jog}).
Furthermore, the high velocities of colliding gas during mergers might act
as dynamical heating that counteracts a fast cooling which would prevent
an efficient cluster formation. In this way metal-rich gas, where the
cooling times are normally short, could also efficiently form GCs.\\ 

What is the observed cluster formation efficiency in merger and 
starburst galaxies? Meurer et al.
(\cite{meur}) investigated the ultraviolet (UV) properties of young
clusters in nine starburst galaxies (blue compact dwarfs as well as
ultraluminous mergers). On average, about 20\% of the UV luminosity
comes from clusters. But is this percentage sufficient to increase the specific
frequency of the GCS? Before answering this question one has to know how many
young clusters will survive the evolution of several Gyr.

Fritze v.~Alvensleben \& Kurth (\cite{frit97}) calculated with the help of 
stellar population evolutionary models (Fritze v.~Alvensleben \& Burkert
\cite{frit95}) that the young Antennae (NGC 7252) clusters will evolve into
a typical GCS. However, they do not exclude the possibility that up to 60\% 
of the
present clusters may be destroyed by dynamical effects during the evolution 
of the cluster system.
This value is the result of semi-analytical model calculations by Vesperini 
(\cite{vesp}), who simulated the evolution of a original GC population in a 
spiral. Note that most of the destroyed clusters are low mass clusters.
Thus the destroyed cluster mass is a much smaller
percentage of the initial total cluster mass. On the other hand,
Okazaki \& Tosa (\cite{okaz95}) estimated that about 60\% in mass
of an initial GC population, whose initial mass function $\phi$ is approximated
by the power law $\phi = {\rm d}N/{\rm d}M \propto M^{-\alpha}$, with $\alpha
\simeq 2$, will be destroyed after evolving into the present GCLF.
In the following we will assume that a dissolution of 20 to 60\% of the 
cluster mass or light is reasonable.

\subsection{Increase of $S_N$ in a starburst galaxy}

Coming back to the question, whether a strong starburst like those investigated
by Meurer et al. (\cite{meur}) can increase $S_N$,
we make a simple calculation: we start with a gas-rich dwarf that has an
absolute luminosity of $M_V = -16.0$ mag and 5 GCs, which means
$S_N = 2$ (typical for spirals, e.g. Zepf \& Ashman \cite{zepf93a}).
We assume that a starburst occurs which involves 10\% of the total mass, of
which 20\% will be transformed into clusters.
For the duration of the burst this increases $M_V$ of the galaxy
to about $-17.8$ mag (assuming that the young stellar population is about
4 mag brighter than a faded old one, Fritze v.~Alvensleben \& Burkert 
\cite{frit95}). About 12 Gyr after the burst $M_V$ has faded again to
$-16.1$ mag. At this time the total luminosity of the clusters is
$M_V = -11.8$ if no cluster has been destroyed, or $M_V = -11.0$ if 
as many clusters as corresponding to about 50\% of the total cluster light 
have been destroyed.
Adopting for the evolved GCS a typical GCLF (t5-function) with a
turnover magnitude of $M_{\rm V,TO} = -7.4$ mag and a dispersion of
$\sigma = 1.0$ (e.g. Kohle et al. \cite{kohl}), about 30 or 18 GCs have 
survived, respectively. The ``specific frequency'' of the GCS at the
time of the starburst is still quite low, $S_N = 2.7$ for 35 clusters (or
cluster candidates), since the galaxy itself is dominated by the young bright
stellar light. Such low $S_N$ values were determined for Local Group irregulars
including the LMC (Harris \cite{harr91a}). After 12 Gyr $S_N$ has 
increased 
significantly, $S_N = 8.4$ for 23 GCs (or even $S_N = 12.7$ for 35 GCs).

If the starburst is 10 times weaker (1\% of the total mass),
only 3--5 clusters would have survived and the resulting $S_N$
is only slightly larger than before, $S_N = 3.5 \pm 0.5$.\\

\subsection{Estimation of the final $S_N$ of a starburst}

Furthermore we want to answer the following question. What is the
specific frequency
of the starburst itself, without an already existing old stellar population?
In other words, we consider an isolated gas cloud and assume that some mechanism
has triggered a starburst as strong as observed in starburst galaxies. Then we
``destroy'' about
20 to 60\% of the cluster light, and look how many GCs survived compared to
the total luminosity of the whole system. Note that stars and clusters
fade in the same way (according to the models by Fritze v.~Alvensleben \&
Burkert \cite{frit95}). The final GCLF has by definition the shape
of a t5-function with $M_{\rm V,TO} = -7.4$ mag and $\sigma = 1.0$.
We assume that immediately after the burst 20\% of
the light comes from clusters and that 1000 GCs will survive the evolution.
Table~\ref{tsimsf} summarizes the results. In column 1 the fraction
$f_{\rm destr}$ of GC light that has been disrupted during the evolution is
given. Columns 2, 3, and 4 are the absolute luminosities of the evolved
GCs, stars, and the total system, respectively. Column 5 gives the resulting
$S_N$ of the system.

\begin{table}
\caption{\label{tsimsf} Specific frequencies of a starburst, in which 1000
GCs are contained in the final GCLF. The values have been derived under the
assumption that 20\% of the initial starburst light comes from clusters
(Meurer et al. \cite{meur}) and a fraction of the cluster light 
$f_{\rm destr}$ has 
been supplied to the field star light due to cluster destruction
}
\begin{flushleft}
\begin{tabular}{rrrrr}
\hline\noalign{\smallskip}
$f_{\rm destr}$ & $M_{V,\rm GCs}$ & $M_{V,\rm stars}$ & $M_{V,\rm total}$ & 
$S_N$ \\
\noalign{\smallskip}
\hline\noalign{\smallskip}
0.2 & -15.6 & -17.4 & -17.6 & 89.0 \\
0.3 & -15.7 & -17.6 & -17.8 & 75.7 \\
0.4 & -15.7 & -17.9 & -18.0 & 62.3 \\
0.5 & -15.7 & -18.1 & -18.2 & 51.1 \\
0.6 & -15.6 & -18.3 & -18.4 & 44.5 \\
\noalign{\smallskip}
\hline
\end{tabular}
\end{flushleft}
\end{table}

The calculations show that $S_N$ in an isolated starburst is
very high, $40 < S_N < 90$. We note that there exists no evidence that
such a high $S_N$ can be the result of a simple undisturbed galaxy formation.
In particular, dEs, whose structural properties are most easily explained by 
a starburst followed by a supernova-driven wind (e.g. Dekel \& Silk 
\cite{deke}), have much lower $S_N$ values.
However, in the context of the galaxy infall scenario, our calculations
might imply that stripped gas from galaxies -- and especially dwarf
galaxies expell their gas most easily (i.e. Dekel \& Silk \cite{deke}) -- can
significantly increase the GC $S_N$ of the central GCS, if it suffers
a starburst comparable to that observed in starburst galaxies.
In Sect.~9 we apply these results to NGC 1399.
%
%%%%%%%%%%%%%%%%%%%%%%%%%%%%%%%%%%%%%%%%%%%%%%%%%%%%%%%%%%%%%%%%%%%%%%%%%%%
%%%%%%%%%%%%%%%%%%%%%%%%%%%%%%%%%%%%%%%%%%%%%%%%%%%%%%%%%%%%%%%%%%%%%%%%%%%
%%%%%%%%%%%%%%%%%%%%%%%%%%%%%%%%%%%%%%%%%%%%%%%%%%%%%%%%%%%%%%%%%%%%%%%%%%%
%

\section{Constraints and estimations for the dwarf galaxy infall scenario
in the Fornax cluster}

\subsection{Why the present day dwarf galaxy population supports the hypothesis
of early infall}

In a scenario where a sufficient number of dwarfs has been dissolved into a
cD halo, one would expect a flat faint end slope of the LF
compared to the initial value. L\'opez-Cruz et al. (\cite{lope}) 
found in a sample of 45 clusters that clusters with a pronounced cD galaxy
indeed tend to have a flat LF faint end slope.
This is what we also find for the Fornax cluster.

Furthermore, the surface density slope of
dE and dS0 galaxies within the core radius of the cluster ($r_c = 0\fdg7$)
is flatter than the slope of all possibly dissolved and/or stripped material:
cD halo stars, GCs, and perhaps rest gas. 
This is consistent with White's (\cite{whit87}) argument that disrupted 
material is more concentrated than the relaxed galaxy distribution.

Moreover, like in other evolved clusters, the gas-rich
late-type Fornax galaxies are found at the outskirts of the cluster, whereas
the early-type (possibly stripped) dwarfs are more concentrated towards the
center (Ferguson \cite{ferg89}). 

One also expects that the LF of compact dwarfs is steeper than the LF
of less compact dwarfs (since they are more easily disrupted). Indeed, 
in the Fornax cluster the LF of the non-nucleated dE/dS0s is flatter than the
LF of the (on the average) more luminous nucleated dE/dS0s.

Furthermore, if the stripping of gas and stars was more effective in the inner
regions, one would expect to find a larger number of fainter remnants in the 
center than outside. This is indeed seen for the non-nucleated dE/dS0s. 
The fainter dwarfs are more concentrated to the center than the brighter ones
(Ferguson \& Sandage \cite{ferg88}).

Finally, two probable candidates for survived
nuclei of dissolved dwarfs have been found (see Paper~1 and 2), which would
indicate that also dwarfs from the brighter nucleated dE/dS0 population have 
been dissolved.

\subsection{Constraints from the metallicity distribution}

Constraints on the metallicity have to be considered, if one assumes that
the metallicity distribution of the GCs is bimodal rather than equally 
distributed
over the range of possible GCs metallicities, $-2.0 < [Fe/H] < 0.0$ dex.
Primordial gas, expelled and stripped from low mass dwarfs, is normally
believed to be a 
contributor to the metal-poor GC subpopulation ($[Fe/H] < -1.1$ dex), if 
transformed into GCs. However, this may not always be the case.
Mac Low \& Ferrara (\cite{macl}) calculated that
galaxies less massive than about $10^8$ M$_{\sun}$ can eject metals 
from supernovae into the intergalactic medium easier than their interstellar 
gas. Thus, the
expelled gas might also be enriched from the supernovae ejecta and more
metal-rich GCs could have been formed as well.
The capture of GCs of early-type dwarf galaxies 
can only have contributed to the metal-poor GCs, since all
observed GCs in such dwarfs
seem to be more metal-poor than $[Fe/H] = -1.2$ dex (e.g. Minniti et al. 
\cite{minn96b}). C\^ot\'e et al. (\cite{cote}) have shown via Monte Carlo
simulations that the capture of dwarf galaxies can indeed reproduce the bimodal
color distribution around M49 and M87 under the assumption that the red
globular cluster population is the intrinsic GCS of the galaxies. 
The mean metallicity of the captured GCs peaks around $\overline{[Fe/H]}
= -1.3$ dex for a steep initial LF slope in their simulations. However, they
do not include dwarf galaxies fainter than $M_V = -13$ mag. As shown in
Sect.~6, the inclusion of these galaxies and an extrapolation of the 
metallicity--luminosity relation for their GCs, can push the mean metallicity
of the captured GCs to a lower value.

Since the metallicity of the accreted stellar population of dwarfs
itself is between $-2.0$ to $-0.6$ (taking the values of Local Group dwarfs,
e.g. Grebel \cite{greb97}), one should also see a low metallicity in the 
cD halo light.
Unfortunately, the metallicity determination of the cD halo is quite
difficult due to the low surface brightness, and in the center the light
of the bulge of NGC 1399 would dominate a metal-poor halo
component. On the other hand, the metallicity of the halo is probably a 
mixture
of different metallicities, if one assumes that not only metal-poor dwarfs
have been dissolved, but also stellar
populations of more massive galaxies had been stripped and new stars from
infalling gas of higher metallicities might have formed.

To explain by the accretion scenario the majority of the red metal-rich GC 
subpopulation ($[Fe/H] \simeq -0.6$ dex), either
already existing GCs of this metallicity had to be captured, or GCs had to be
newly formed from enriched infalling gas. This is possible, if one allows that
the gas-rich dwarfs first had time to enrich their interstellar matter to at 
least $-0.8$ dex, before the stripping of the gas became important and/or
before new cluster formation in these dwarfs has been triggered by interaction
processes. 
In the LMC, for example, no clusters were formed between about 3 to 10
Gyr ago. Whereas the few old clusters have a metallicity of about $-1.8$ dex,
the younger clusters have metallicities around $-0.4$ dex (e.g. Olszewski et al.
\cite{olsz91b}, Hilker et al. \cite{hilk95}). Concerning the time scale 
for metal enrichment in spirals, e.g. M\"oller et al. (\cite{moel})
estimated that 2-3 Gyr is enough time to enrich the iron abundance of the
interstellar medium from $-1.5$ dex to about $-0.6$ dex for early-type spirals
(Sa, Sb), whereas at least 7 Gyr are needed for late-type spirals (Sc, Sd).
Similarly,
Fritze-v.~Alvensleben \& Gerhard (\cite{frit94}) calculated the metallicity of 
a secondary GC population in a early-type spiral-spiral merger to be about
$[Fe/H] = -0.6$ dex after about 2 Gyr of their life time, and after about 
8 Gyr in a late-type spiral-spiral merger. 

Assuming that the metal-rich GC population in Fornax was formed by the
infall of all types of gas-rich galaxies, one therefore should expect a
range of ages inbetween them in order to account for a metallicity peak 
around $-0.6$ dex. First, the metal-rich GCs should be at least 2 Gyr 
younger than the metal-poor GCs with $[Fe/H] = -1.3$ dex. Second, their age 
spread should be in the order of 2-6 Gyr.
It would be interesting to know whether this prediction can be proved or 
disproved in further investigations.

We note that a large number of metal-rich GCs also might have been accumulated
by stripping from more massive early-type galaxies. According to the 
metallicity--luminosity relation by C\^ot\'e et al. (\cite{cote}), the mean 
metallicity, $[Fe/H] = -0.6$ dex, of the red GC population around NGC~1399
corresponds to a luminosity of the former parent galaxies of about $M_V = -20$
mag. This value is typical for the low-luminosity ellipticals in Fornax,
as for example NGC~1374, NGC~1379, and NGC~1427.

\subsection{Constraints from the spatial distribution of globular clusters}

A further point that has to be explained is why the red GC population is more 
concentrated than the blue one (Forbes et al. \cite{forb97}). The answer 
might be that star formation from 
infalling gas is more concentrated to the inner part of the dense cluster 
core, as it is also expected for a merging scenario (Ashman \& Zepf 
\cite{ashm92}).
Another possibility is that most of the red GCs have nothing to do with
a secondary formation or accretion process, but rather belong to the
original GC population of the bulge of NGC~1399. However, as we show in the
next section, not more than about 1300 GCs can belong to the bulge light, 
if one assumes reasonable values of the initial GC specific frequency.
This comprises only about half of the red GC subpopulation.

\section{The correct mix of accreted and newly formed GCs}

Let us imagine that the infall of dwarf galaxies and gas was really the 
dominating process for the building-up of the cD halo and the GCS.
How many dwarfs and their transformed gas would then have contributed to the 
cD halo light and how many GCs might belong to the cD halo?

\begin{table*}
\caption{\label{tmixgc} Possible mixtures of different processes of GC 
accretion,
formation, and stripping that can explain the observed cD halo properties.
In each line, the contribution of each process can be added to the following
properties of the cD halo: $M_{V,cD} = -21.65$ mag, $N_{tot,cD} = 4500$, and
$S_N = 10$. See text for more details.
}
\begin{flushleft}
\begin{tabular}{crrrrrrrrr}
\hline\noalign{\smallskip}
 & \multicolumn{3}{c}{accretion of GCs} & 
\multicolumn{3}{c}{new formation of GCs} & \multicolumn{3}{c}{stripping of GCs}\\
case & $M_V$ & $S_N$ & $N_{\rm GC}$ & $M_V$ & $S_N$ & $N_{\rm GC}$ 
& $M_V$ & $S_N$ & $N_{\rm GC}$\\
\noalign{\smallskip}
\hline\noalign{\smallskip}
1 & -21.11 & 9.0 & 2500 & -20.63 & 11.2 & 2000 & & & \\
2 & -21.11 & 9.0 & 2500 & -20.53 &  9.2 & 1500 & -18.05 & 30 & 500\\
3 & -21.52 & 6.2 & 2500 & -19.25 & 40.0 & 2000 & & & \\
4 & -21.59 & 5.8 & 2500 & -18.49 & 80.0 & 2000 & & & \\
5 & -21.52 & 4.9 & 2000 & -18.81 & 60.0 & 2000 & -18.05 & 30 & 500\\
6 & -21.22 & 4.9 & 1500 & -20.44 & 20.0 & 3000 & & & \\
7 & -21.45 & 3.9 & 1500 & -19.69 & 40.0 & 3000 & & & \\
8 & -21.16 & 3.4 & 1000 & -20.44 & 20.0 & 3000 & -18.05 & 30 & 500\\
\noalign{\smallskip}
\hline
\end{tabular}
\end{flushleft}
\end{table*}

NGC 1399 possesses about 5800 globular clusters (see Sect.~3.2).
About 1300 of them would belong to the bulge,
$M_{\rm V,gal} = -21.5$ mag (see Sect.~3.3.1), if one assumes an
initial specific frequency of $S_N = 3.2$, which is the mean value for the
other ellipticals in the Fornax cluster, except NGC 1404 and NGC 1380.
That means that 4500 GCs would belong to the cD halo and its specific frequency
would be about $S_N = 10\pm1$. Note that half of the total GCS ($= 2900$ GCs)
are assigned
to the metal-poor peak around [Fe/H] $\simeq$ $-$1.3 dex, and
therefore at least 1600 metal-rich GCs ([Fe/H] $\simeq$ $-$0.6 dex) have to
be explained by the infall scenario,
if one assumes that all 1300 remaining bulge GCs belong to the metal-rich
sub-population.

How can dwarf galaxies account for such a high $S_N$?

As presented in Sect.~5, there are mainly three scenarios possible.
Firstly, accreted gas-poor dwarfs possessed high GC frequencies themselves.
In this case, the average $S_N$ of all accreted dwarfs and GCs can have values
between 4 and 22 depending on the initial conditions (see Table 5).
Secondly, the infalling gas of previously gas-rich dwarfs was effectively
converted into globular clusters. Regarding the starburst as
an isolated entity its resulting systems of stars and clusters can have $S_N$
values between 40 and 90 (see Table 6).
Finally, the stripping of GCs from dwarf galaxies was more effective than
the stripping of their field population. That this is in principal possible
is indicated by the fact that the $S_N$ value of the outer parts of 
galaxies that are primarily affected by stripping can be in the order of 
30 (see Sect.~5.1, case 1b).

Among these 3 possibilities the stripping of GCs from dwarf galaxies
most probably plays a minor role. Even if all 50 dE/dS0s within
the core radius of the galaxy distribution are remnants, whose outer GCs
have been stripped off, we calculate that maximally some hundred GCs have
been captured by this process, assuming an initial $S_N = 5.5$, $S_N = 30$
for the stripped stars and GCs, and a final $S_N = 3.0$ for the remnant
(similar to the values for NGC~4472, McLaughlin et al. \cite{mcla94b}). In the
following, we consider the case that at most 500 GCs have been stripped.\\

What is the correct mixture of the two other processes that fulfill the
following assumptions?\\

(1) The cD halo has been formed only by accreted and newly formed matter, 
  and its total luminosity is $M_{\rm V,cD} = -21.65$ mag

(2) The specific frequency of the accreted and newly formed GCs with respect
  to the halo luminosity is $S_N = 10$, ($= 4500$ GCs)

(3) The 4500 GCs in the cD halo consist of 2500 metal-poor (blue) and 2000 
   metal-rich (red) GCs
  (this implies that the GCS of the bulge has 400 metal-poor and
  900 metal-rich GCs)

(4) GCs captured by accretion and stripping of dwarfs can only be metal-poor\\

In Table~\ref{tmixgc} we present the possible mixtures of the 3 processes,
starting with cases for which GC accretion is dominant and ending with
cases in which most GCs have been formed from infalling gas.

In the first five cases we assumed that all metal-poor GCs were captured 
or stripped.
Assuming a high $S_N$ value for the accretion process ($S_N = 9$, cases 1 and 
2), the cluster formation efficiency (CFE) does not need to be as high as
estimated for merger and starburst situations (see Table~\ref{tsimsf}).
However, as discussed in Sect.~6,
a high $S_N$ requires a high accretion rate of dwarf galaxies, a steep
initial slope of the faint end of the galaxy LF, and also very faint dwarf
galaxies should have possessed at least one GC. The faintest dwarf galaxies with
a GCS observed so far are the Local Group dSphs Fornax and Sagittarius
($M_V \simeq -12.5$ mag). 

The other way around, if starbursts
from stripped gas can produce a high $S_N$ value ($40 < S_N < 80$, cases 3-5)
and have formed 2000 metal-rich GCs, the $S_N$ for the accreted metal-poor GCs
is in the order 5-6. Such values can easily be achieved by the accretion
scenario presented in Sect.~6 under various reasonable initial conditions 
(see Table 5).

In the cases 6 to 8 we assumed that the majority of the GCs had their origin 
from infalling gas with a low value of the estimated starburst CFEs
($20 < S_N < 40$). The specific frequency for the remaining 1500 accreted GCs
then can be very low ($3 < S_N < 5$), very faint dwarfs do not need to
possess GCs, and the numbers of dissolved dwarfs can be in 
the order of 250-500. However, one then has to assume that GCs formed 
from metal-rich as well as metal-poor gas and that most of the original
dwarf galaxies have been very gas-rich.

\section{Concluding remarks}

We have summarized the properties of the different components of the central 
galaxy 
NGC~1399 (GCS, cD halo, bulge) and the dwarf galaxy population in the center 
of the cluster. We have analysed under which circumstances
the GCS and cD halo can be explained by the infall and accretion of gas-poor as
well as gas-rich dwarf galaxies.

Estimations of the GC formation efficiency from infalling gas,
and simulations of the accretion of GCSs from early-type dwarfs have shown
that the building-up of the cD halo and central GCS by dwarf galaxy and gas 
infall alone is only possible under special conditions during the
formation and evolution of the cluster. Depending on the leading process
which contributed most to the rich GCS, the following conditions are required
to fulfill the observed properties:
\begin{itemize}
\item There are as many blue (metal-poor) as red (metal-rich) GCs seen
around NGC~1399. Since not all metal-rich GCs can be assigned to the bulge of 
NGC~1399 when adopting a reasonable $S_N$, the formation of secondary 
metal-rich GCs from stripped gas (or within the dwarf galaxies) probably was
an effective process besides the capture of metal-poor GCs.
\item The stripping and capture of GCS of gas-poor dwarf galaxies can only 
account for the metal-poor GC population.
\item If the accretion of gas-poor dwarfs was
a dominating process, the faint end slope of their initial LF had to be as
steep ($\alpha < -1.4$) as it is predicted in CDM models in order to 
provide 
a sufficient number of dwarfs that have been disrupted in the central galaxy.
\item A steep faint end slope of the initial LF leads to a mean metallicity
of the captured dwarfs that is about 0.5 dex more metal-poor than the observed
value of the blue GC population around NGC 1399.
\item Furthermore, in the accretion dominated scenario, an unlikely high 
number of dwarfs 
($\simeq 6000$, see Table 5) had to
be accreted, and about 50\% of the fainter dwarfs ($-12.5 < M_V < -10.5$)
must have possessed at least one GC in order to produce high $S_N$ values.
\item A very efficient increase in $S_N$ of the central GCS by the formation 
of GCs from gas can be achieved, if the the cluster formation efficiency
was as high as in merging or starburst galaxies.
\item If the majority of the GCs (metal-poor as well as metal-rich ones)
formed from stripped gas, a significant fraction
of the gas was enriched to at least $-1.0$ dex before forming
GCs in order to explain the bimodal metallicity distribution of the central GCS.
This implies that the metal-rich GCs should be at least 2 Gyr older than 
the metal-poor ones and should show a significant age spread.
\end{itemize}

Certainly, some of the requirements are quite restrictive. We conclude that
the infall of dwarf galaxies can principally explain many properties in
the center of the Fornax cluster, but is most probably not the only process
that has been active. Certainly, also the brighter, more massive galaxies
were envolved by the interaction processes in the central region
of the Fornax cluster. 
A natural extension of the dwarf galaxy infall scenario is, for example, 
the stripping (and early merging) of giant galaxies, ellipticals and 
spirals (as mentioned in Sect. 8.2). Besides the low-luminosity ellipticals
in Fornax, very likely candidates for stripping are the central giant
galaxies NGC~1380 and NGC~1404, which have low GC
specific frequencies, and might therefore have provided a significant fraction
of the central GCS (see Kissler-Patig et al. \cite{kiss99a}).

We are aware of the fact that our proposed scenario has to be tested and
confirmed by further theoretical as well as observational work.
Especially, it has to be shown in N-body simulations whether the accretion
rate of dwarf galaxies can be very high, and what is the dynamical behaviour
of stripped and accreted GCs in the central cluster potential.
Furthermore, it has to be tested under which conditions a high
cluster formation efficiency can be obtained from stripped gas (whether it 
is comparable
to a starburst situation in a galaxy or not).
On the observational side it has to be shown, whether the faintest dwarf
galaxies possess GCs or not. Further investigation of the faint end slopes 
of galaxy LFs for clusters with very different properties (redshift, richness,
compactness, existence of a cD galaxy, etc.) will show whether the 
proposed scenario is compatible with the findings. Observations more
sensitive to the ages of GCs (i.e. measurement of line indices) will prove
or disprove the predictions of an age spread among the GCs.

%
%%%%%%%%%%%%%%%%%%%%%%%%%%%%%%%%%%%%%%%%%%%%%%%%%%%%%%%%%%%%%%%%%%%%%%%%%%%
%%%%%%%%%%%%%%%%%%%%%%%%%%%%%%%%%%%%%%%%%%%%%%%%%%%%%%%%%%%%%%%%%%%%%%%%%%%
%%%%%%%%%%%%%%%%%%%%%%%%%%%%%%%%%%%%%%%%%%%%%%%%%%%%%%%%%%%%%%%%%%%%%%%%%%%

\acknowledgements
This research was partly supported by the DFG through the Graduiertenkolleg
`The Magellanic System and other dwarf galaxies' and through
grant Ri 418/5-1 and Ri 418/5-2. MH thanks Fondecyt Chile for support
through `Proyecto FONDECYT 3980032' and LI for support through `Proyecto 
FONDECYT 8970009'.

\enddocument